\documentclass{article} %
\usepackage{iclr2025_conference,times}

\usepackage{amsmath,amsfonts,bm}

\def\eqref#1{equation~\ref{#1}}

\def\1{\bm{1}}

\DeclareMathAlphabet{\mathsfit}{\encodingdefault}{\sfdefault}{m}{sl}
\SetMathAlphabet{\mathsfit}{bold}{\encodingdefault}{\sfdefault}{bx}{n}

\usepackage[hyphens]{url}
 \makeatletter
 \def\url@leostyle{%
   \@ifundefined{selectfont}{\def\UrlFont{}}%
   {\def\UrlFont{}}%
 }
 \makeatother
\definecolor{darkred}{RGB}{90, 0, 0}
\definecolor{darkblue}{RGB}{0,0,130}
\usepackage[bookmarks=false, colorlinks=true, plainpages=false,  linkcolor=darkred, citecolor=darkred, urlcolor=darkblue, filecolor=blue]{hyperref}

\renewcommand{\footnotesize}{\fontsize{8pt}{9pt}\selectfont}

\usepackage{paralist}
\usepackage{enumitem}

\usepackage{graphicx}	%
	\setkeys{Gin}{width=\textwidth, height=\textheight, keepaspectratio}
	\usepackage[skip=0pt,font=footnotesize,justification=centering]{subcaption}

\newif\ifcomment
\commenttrue

\usepackage[hang,flushmargin,stable]{footmisc}

\renewcommand{\footnoterule}{%
  \kern -3pt
  \hrule width 1in
  \kern 2pt
}
\usepackage[compact]{titlesec}
\titlespacing*{\section}{0pt}{*1}{1pt}
\titlespacing*{\subsection}{0pt}{*1}{1pt}

\setitemize{noitemsep,topsep=0pt,parsep=0pt,partopsep=0pt,leftmargin=2em}

\ifcomment
	\newcommand{\edc}[1]{\textbf{\em\color{red}EDC: #1}}
	\newcommand{\gvg}[1]{\textbf{\em\color{blue}GG: #1}}
	\newcommand{\sundar}[1]{\textbf{\em\color{purple}MS: #1}}
	\newcommand{\sof}[1]{\textbf{\em\color{brown}SOF: #1}}
\else
	\newcommand\edc[1]{}
	\newcommand\gvg[1]{}
	\newcommand\sundar[1]{}
	\newcommand\sof[1]{}
\fi

\newcommand{\descr}[1]{\noindent\textbf{#1}}

\title{Understanding the Impact of Data Domain \\ Extraction on Synthetic Data Privacy}

\author{%
  Georgi Ganev$^{1,2}$ \quad Meenatchi Sundaram Muthu Selva Annamalai$^{1}$ \\
  \textbf{Sofiane Mahiou}$^{2}$ \quad \textbf{Emiliano De Cristofaro}$^{3}$ \\
  $^1$UCL \quad $^2$SAS \quad $^3$UC Riverside \\
  \texttt{\href{mailto:georgi.ganev.16@ucl.ac.uk}{georgi.ganev.16@ucl.ac.uk}}
}

\iclrfinalcopy %
\begin{document}

\maketitle

\vspace*{-0.4cm}
\begin{abstract}

Privacy attacks, particularly membership inference attacks (MIAs), are widely used to assess the privacy of generative models for tabular synthetic data, including those with Differential Privacy (DP) guarantees.
These attacks often exploit outliers, which are especially vulnerable due to their position at the boundaries of the data domain (e.g., at the minimum and maximum values).
However, the role of data domain extraction in generative models and its impact on privacy attacks have been overlooked.
In this paper, we examine three strategies for defining the data domain: assuming it is externally provided (ideally from public data), extracting it directly from the input data, and extracting it with DP mechanisms.
While common in popular implementations and libraries, we show that the second approach breaks end-to-end DP guarantees and leaves models vulnerable.
While using a provided domain (if representative) is preferable, extracting it with DP can also defend against popular MIAs, even at high privacy budgets.
\end{abstract}

\section{Introduction}
Differentially Private (DP) synthetic tabular data promises to support the safe release of sensitive data by training generative machine learning models while limiting individual-level information leakage.
This approach is gaining significant traction~\citep{jordon2022synthetic, hu2024sok,cristofaro2024synthetic}
and is increasingly being deployed in real-world applications, from public releases of census data~\citep{nasem2020census, ons2023synthesising, hod2024differentially} to data sharing in financial and healthcare contexts~\citep{ico2023synthetic, microsoft2022iom}.
Synthetic data has also attracted interest from regulators~\citep{ico2023privacy,ico2023synthetic, fca2024using}, who are shifting focus from assessing the anonymity of released datasets~\citep{eu2014opinion} to evaluating generative models%
~\citep{eu2024opinion}.

In this context, membership inference attacks (MIAs)~\citep{shokri2017membership, hayes2019logan}, %
are used as a measuring stick for privacy leakage.
MIAs are typically evaluated using a privacy game that entails identifying (or crafting) a vulnerable record, training a generative model with it and without it, generating synthetic data, and having the adversary distinguish whether or not that record was used to train the model.
In this game, outliers located at the boundaries of the data domain (e.g., at each column's min or max values) are particularly vulnerable~\citep{stadler2022synthetic, annamalai2024what}. %
However, adding/removing outliers can significantly impact the training of DP generative models, especially the initial pre-processing steps (e.g., scaling, normalization, discretization, encoding, etc.) common in DP synthetic tabular data algorithms.
This presents a unique challenge for tabular data, unlike, e.g., for images or text, where input pixels and tokens have clearly defined domains (e.g., $[0, 255]$ or ASCII characters).

Nonetheless, many implementations and libraries for DP synthetic tabular data have overlooked this issue, as they directly extract data domain from the input data~\citep{zhang2017privbayes, ping2017datasynthesizer, vietri2020new, mckenna2021winning, qian2023synthcity, mahiou2022dpart, du2024towards}.
In this paper, we examine how different strategies for extracting the data domain affect the privacy of DP generative models for tabular synthetic data.
Specifically, we compare a publicly available data domain to extracting the domain directly from the input data---denoted, respectively, as {\em provided} and {\em extracted} domain.
We do so both with and without DP and for two generative models, PrivBayes~\citep{zhang2017privbayes} and MST~\citep{mckenna2021winning}.
Since both models require discretized data, we adapt and assess four DP discretization strategies: uniform, quantile, k-means, and PrivTree~\citep{zhang2016privtree}.
For the MIA, we use the GroundHog attack~\citep{stadler2022synthetic}.

In short, our experiments show that:
\begin{itemize}
	\item Extracting the data domain directly from the input data, %
	which is the common practice, breaks the end-to-end DP guarantees of generative models and exposes outliers to MIAs. %
	\item Assuming that a representative data domain is provided and extracting it with DP (up to $\epsilon=100$) successfully protects outliers from specific MIAs.
  In particular, adopting a DP domain extraction strategy could address many previously identified DP vulnerabilities in open-source implementations and libraries.
	\item The GroundHog attack~\citep{stadler2022synthetic} may be more effective at detecting issues with data domain extraction than with vulnerabilities of the generative models themselves.
\end{itemize}

\descr{From Domain Extraction to Discretization.}
In separate work~\citep{ganev2025importance}, we examine the broader question of discretization in end-to-end DP generative models, primarily focusing on utility.
In contrast, this paper focuses specifically on data domain extraction strategies and their privacy implications, which is closely related to Research Question~4 in~\citep{ganev2025importance}.

\section{Experimental Framework}
\label{sec:fram}
As mentioned, we aim to evaluate the impact of the domain extraction strategy on privacy leakage in (end-to-end) DP generative models using an MIA.
We experiment with three strategies:
1) assuming a provided data domain set to the full dataset's range, regardless of the target record's inclusion/exclusion,
2) extracting it directly from the input data (without DP), as done in numerous publicly available implementations and libraries~\citep{zhang2017privbayes, ping2017datasynthesizer, vietri2020new, mckenna2021winning, qian2023synthcity, mahiou2022dpart, du2024towards}, or
3) extracting the data domain with DP~\citep{desfontaines2020lowering}.

\descr{MIA Instantiation.}
To evaluate the privacy of the resulting synthetic data, we use GroundHog~\citep{stadler2022synthetic}, one of the most widely used MIAs for synthetic tabular data, %
on the Wine dataset~\citep{dua2017data}.
First, we select a vulnerable record as the target, picking the data point furthest from all others in the training set~\citep{meeus2023achilles} and ensuring it lies outside their domain.
Then, we train two sets of 200 shadow models: %
one trained on the full dataset, including the target record, and the other excluding it.
To do so, for each model, we extract the domain, discretize the data, and train the generative model.
We generate synthetic datasets and extract statistical features (minimum, maximum, mean, median, and standard deviation, corresponding to the naive feature set $F_{naive}$ in~\citep{stadler2022synthetic}) from each column of the synthetic datasets.
Half of these datasets are used to train a classifier, and the adversary's success in distinguishing between the two scenarios is measured using Area Under the Curve (AUC), reported on the remaining data.

\descr{Settings.}
We choose PrivBayes~\citep{zhang2017privbayes} and MST~\citep{mckenna2021winning} as our DP generative models, using $\epsilon=1$ for pre-processing (split evenly between domain extraction, when applicable, and discretization) and $\epsilon=1$ for the model (with $\delta=1\mathrm{e}{-5}$ for MST).
We use 20 bins for all discretization strategies and the default hyperparameters for both models.
The selected target record represents a worst-case scenario, consistent with prior work~\citep{stadler2022synthetic, annamalai2024what}, %
given two columns with values significantly larger than for the remaining records (289 and 440 vs.~146.5 and 366.5).
Due to space limitations, we defer additional details, including the DP data extraction method, discretization strategies, DP generative models, and dataset, to Appendix~\ref{app:elems}.

\section{Experimental Evaluation}
\label{sec:exp}

\begin{figure*}[t!]
  \centering
  \begin{subfigure}{0.75\linewidth}
    \includegraphics[width=0.99\linewidth]{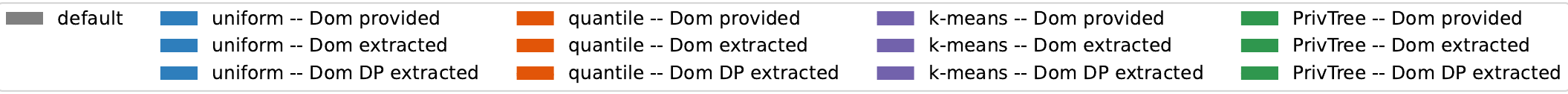}
  \end{subfigure}
  \centering
  \begin{subfigure}{0.75\linewidth}
    \includegraphics[width=0.99\linewidth]{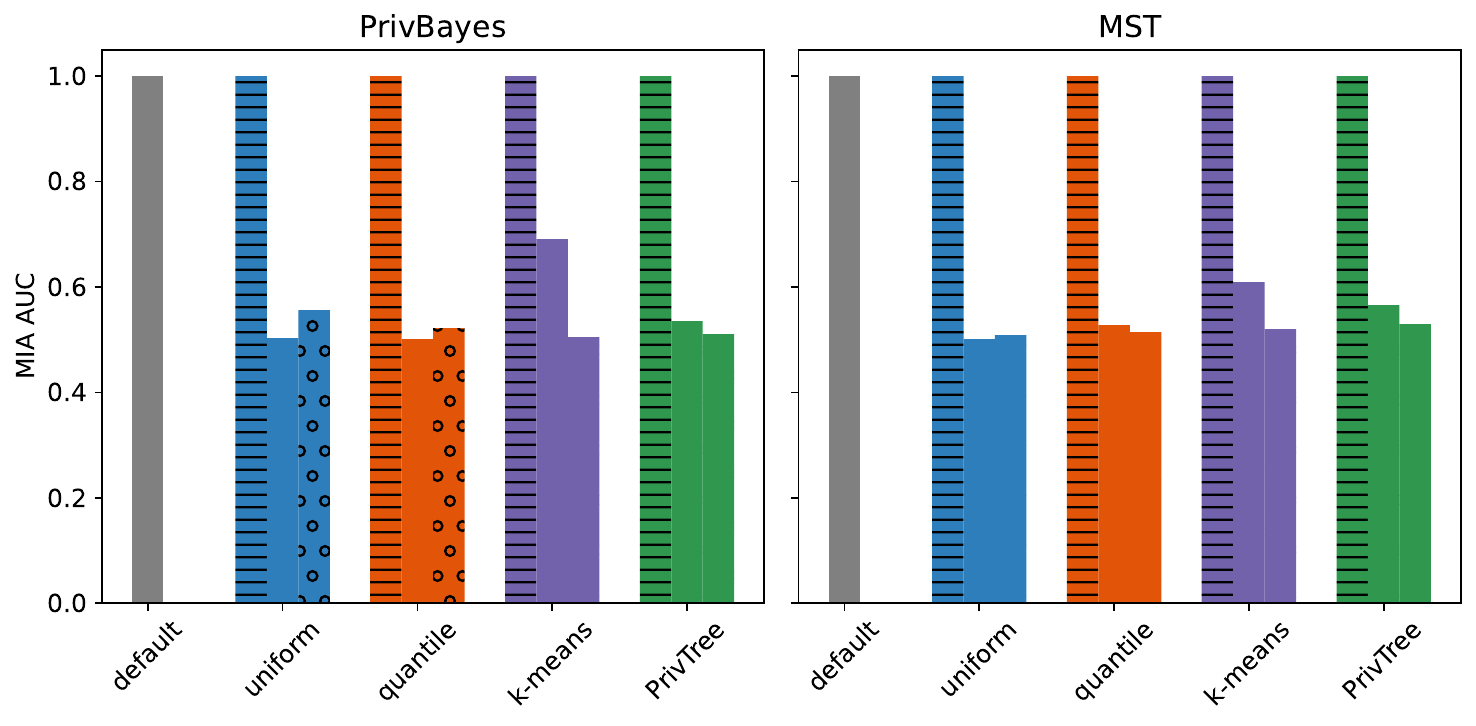}
  \end{subfigure}
  \caption{\small Privacy leakage with \emph{provided} and \emph{extracted} domain (w/ and w/o DP) for the four DP discretizers ($\epsilon=1$) and two DP generative models ($\epsilon=1$) on a target record \emph{outside} the domain of the remaining data.}
  \label{fig:attack_1}
  \vspace{-0.3cm}
\end{figure*}

Figure~\ref{fig:attack_1} provides an overview of our experiments, quantifying the impact of the three domain extraction strategies on privacy leakage as measured by GroundHog~\citep{stadler2022synthetic}'s success with four discretizers and two generative models.

\descr{Direct Domain Extraction.}
Regardless of discretization, extracting the domain directly from the input data (bars with horizontal lines) without a proper privacy mechanism breaks the end-to-end DP guarantees, providing highly informative features.
This enables the adversary to achieve near-perfect accuracy in all cases, rendering the synthetic data non-private regardless of the discretizer or generative model.
This performance is equivalent to that of using the default domain extraction/discretization strategy (grey bars), i.e., uniform discretization with direct domain extraction.

\descr{Provided Domain/DP Domain Extraction.}
By contrast, using methods that respect the end-to-end DP pipeline, i.e., either assuming a provided domain (bars with crosses) or extracting the domain with DP (bars with circles), substantially reduces the adversary's success rate.
With only one exception (when using the k-means discretizer with provided domain), the attack success rate is no better than random guessing. %
This has several important implications, as discussed next.

First, extracting the domain in a DP-compliant way is sufficient to protect against GroundHog~\citep{stadler2022synthetic}'s adversary.
Their success drops significantly even in settings that could be considered non-private, i.e., (discretizer, generator)-$\epsilon$ values of (1, 100), (100, 1), (100, 100), and even (1, 1,000), as shown in Figure~\ref{fig:attack_100} and~\ref{fig:attack_1_1000} (see Appendix~\ref{app:plots}).
The attack becomes effective at higher discretizer $\epsilon$ values, i.e., (1,000, 1) and (1,000, 1,000) (see Figure~\ref{fig:attack_1000_1} and~\ref{fig:attack_1000_1000}); also, recall that it achieves nearly 100\% success when the domain is directly extracted from the data.
This suggests that the effectiveness of the GroundHog attack may primarily be due to domain extraction rather than inherent vulnerabilities in the model.\footnote{These results are specific to GroundHog~\citep{stadler2022synthetic}. Studying the impact of DP domain extraction on other MIAs as well as other models besides PrivBayes and MST, is left to future work.}
To further validate this, we run additional experiments on a target record that is farther away from the others but still within their domain (see Figure~\ref{fig:attack_in} in Appendix~\ref{app:plots}) and observe that the attack's success remains close to random across all $\epsilon$ values.

Second, adopting a DP domain extraction strategy could help address privacy vulnerabilities identified by prior research~\citep{annamalai2024what, ganev2025elusive} in popular model implementations and libraries~\citep{ping2017datasynthesizer, qian2023synthcity} that directly extract the domain from the input data.
In other words, incorporating such techniques could make DP generative model implementations more robust and better align them with end-to-end DP guarantees.

Finally, while extracting the domain in a DP way slightly reduces, on average, the adversary's success compared to using a provided data domain, this may come at the cost of utility.
Therefore, practitioners should prefer using a trusted, provided data domain when available (e.g., codebooks for census data), rather than spending additional privacy budget to extract it.
However, further research is needed to explore these trade-offs and investigate enhanced methods for DP domain extraction.

\section{Conclusion}
This paper focused on an important yet overlooked issue in implementations of DP generative models: how to extract data domain
We show that extracting the data domain directly from the input, which is unfortunately common in the wild~\citep{ping2017datasynthesizer, qian2023synthcity}, breaks DP guarantees and leaves models vulnerable.
We also find that, while using a provided domain (e.g., from public data) is preferable, extracting it with DP can also defend against MIAs, even at large $\epsilon$ values.

Overall, we are confident that our research will shed light on the importance of the integrity of end-to-end DP pipelines when developing and releasing DP generative models.
Our work also highlights the need for further analysis of membership inference attacks against DP generative models, e.g., understanding the extent to which the privacy leakage they exploit may be due to issues like domain extraction rather than inherent vulnerabilities in the models.


\bibliographystyle{iclr2025_conference}

\appendix

\section{More Details on the Experimental Framework}
\label{app:elems}

We now provide details of the experimental framework introduced in Section~\ref{sec:fram}, specifically, DP data extraction methods, DP discretization strategies, DP generative models, and dataset.

\descr{DP Domain Extraction.}
To estimate the domain of numerical data given a privacy budget, $\epsilon$, we implement the algorithm by \citet{desfontaines2020lowering}, also included in the popular OpenDP library~\citep{opendp2021smartnoise}.
It derives bounds using a noisy histogram over an exponential range $[-2^{m}, 2^{m}]$ (with $m$ typically set to 32), determined by iteratively reducing a threshold until at least one bin exceeds it, using the highest and lowest bin edges above the threshold as the domain bounds.

\descr{DP Discretizers.}
We use the following four methods to make the four discretizers satisfy DP (note that the data domain, privacy budget, and number of bins $b$ are provided as input to all discretizers).
For the DP implementation, we use primitives of two well-known open-source libraries, namely, Harvard's OpenDP~\citep{opendp2021smartnoise} and IBM's Diffprivlib~\citep{holohan2019diffprivlib}.

\begin{itemize}
	\item \textit{Uniform} divides the data domain into $b$ intervals of equal width.
  It does not consume any privacy budget and relies solely on the provided data domain to determine bin edges.\smallskip
	\item \textit{Quantile} distributes data such that each bin contains approximately an equal fraction of data points, specifically $1/b$.
  The privacy budget $\epsilon$ is split evenly across a given number of bins, with each quantile calculated using $\epsilon/b$.
  We use the method proposed by~\citet{smith2011privacy}, which samples quantile values from a discrete distribution.
  Each $q_i$ is computed as $(x_{i+1} - x_i) \cdot \exp(-\epsilon |i - \alpha n|)$, where $x_i$ is the value at index $i$ in the sorted dataset, $\alpha$ is the target quantile.\smallskip
	\item \textit{K-means} employs a standard k-means clustering algorithm to group the data into clusters and then splits them into non-overlapping intervals.
  It is based on~\citep{su2016differentially}, which adds Geometric noise~\citep{ghosh2009universally} to the counts of the nearest neighbors for cluster centers and Laplace to the sum of values per dimension.
	However, some clusters may occasionally be empty, resulting in fewer than $b$ bins.\smallskip
	\item \textit{PrivTree}~\citep{zhang2016privtree} is a tree-based method that recursively splits the data domain into subdomains.
  It ensures DP by adding Laplace noise~\citep{dwork2006calibrating} to the count at each step.
  Subdomains are further split if the noisy count exceeds a threshold, $\tau$; otherwise, they form leaves, with bin edges corresponding to the domains of all leaves.
  The threshold parameter $\tau$ is set to ${1}/{b}$, making $b$ an upper limit for the actual number of bins produced.
\end{itemize}

\descr{DP Generative Models.}
The two DP generative models we use, PrivBayes~\citep{zhang2017privbayes} and MST~\citep{mckenna2021winning}, rely on the {\em select–measure–generate} paradigm~\citep{mckenna2021winning}, %
as they: 1) select a collection of (low-dimensional) marginals, 2) measure them privately with a noise-addition mechanism, and 3) generate synthetic data consistent with the measurements.

PrivBayes~\citep{zhang2017privbayes}
uses a Bayesian network to select $k$-degree marginals by optimizing the mutual information between them, using the Exponential mechanism~\citep{dwork2006our}.
Then, propagating though the network, the model relies on the Laplace mechanism~\citep{dwork2006calibrating} to measure noisy counts and translate them to conditional marginals, which could later be sampled to generate synthetic data.
MST~\citep{mckenna2021winning}
forms a maximum spanning tree (an undirected graph) of the underlying correlation graph by selecting all one-way marginals and a collection of two-way marginals.
These marginals are noisily measured via the Gaussian mechanism~\citep{mcsherry2007mechanism}.
Finally, to create new data, the measurements are processed through Private-PGM~\citep{mckenna2019graphical}.

\descr{Wine Dataset}~\citep{dua2017data}. As mentioned, our experiments are run on the Wine dataset, which consists of 4,898 wine samples, each described by 11 continuous physicochemical attributes, with the goal of modeling wine quality.

\section{Additional Plots}
\label{app:plots}

\begin{figure*}[t!]
  \centering
  \begin{subfigure}{0.8\linewidth}
    \includegraphics[width=0.99\linewidth]{plots/legend.pdf}
  \end{subfigure}
  \centering
  \begin{subfigure}{0.3275\linewidth}
    \includegraphics[width=0.99\linewidth]{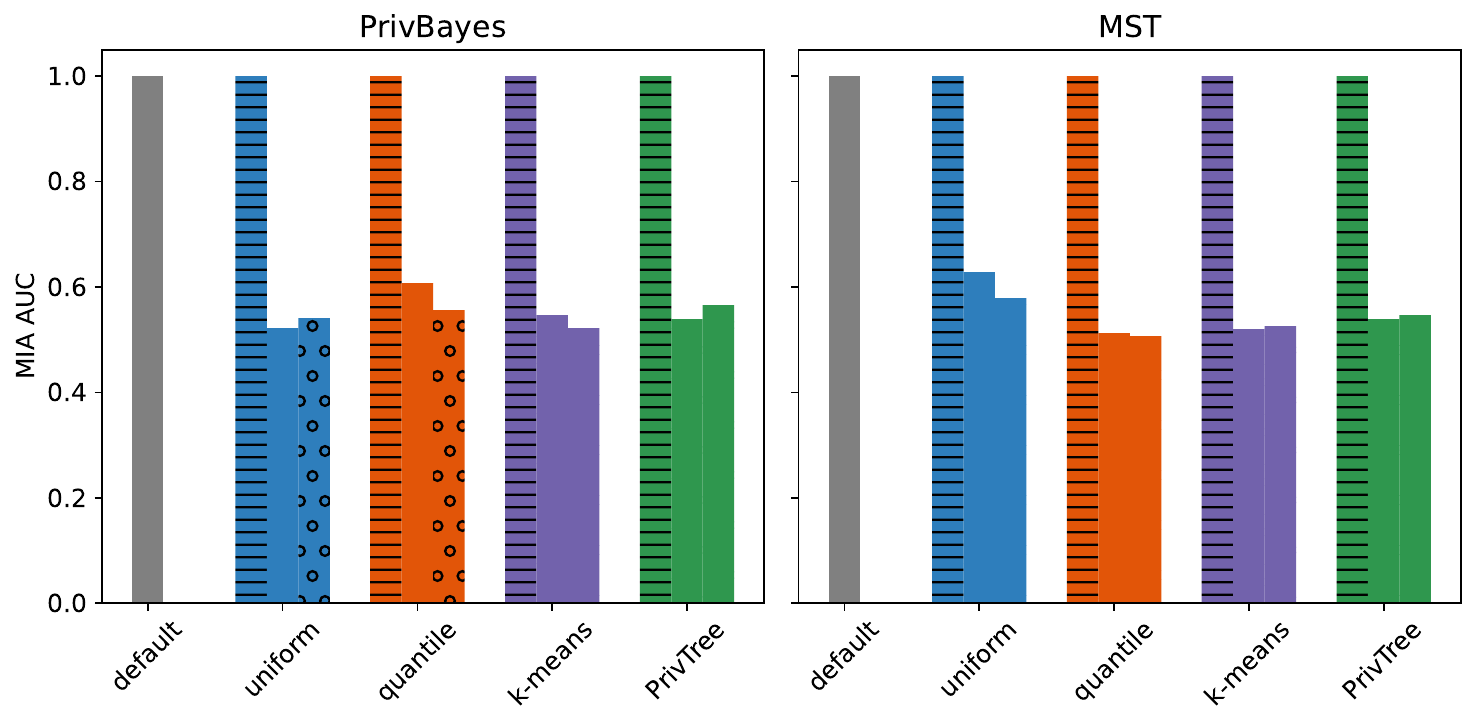}
    \caption{Discretizer ($\epsilon=1$), \\Generator ($\epsilon=100$)}
  \end{subfigure}
\hfill
  \begin{subfigure}{0.3275\linewidth}
    \includegraphics[width=0.99\linewidth]{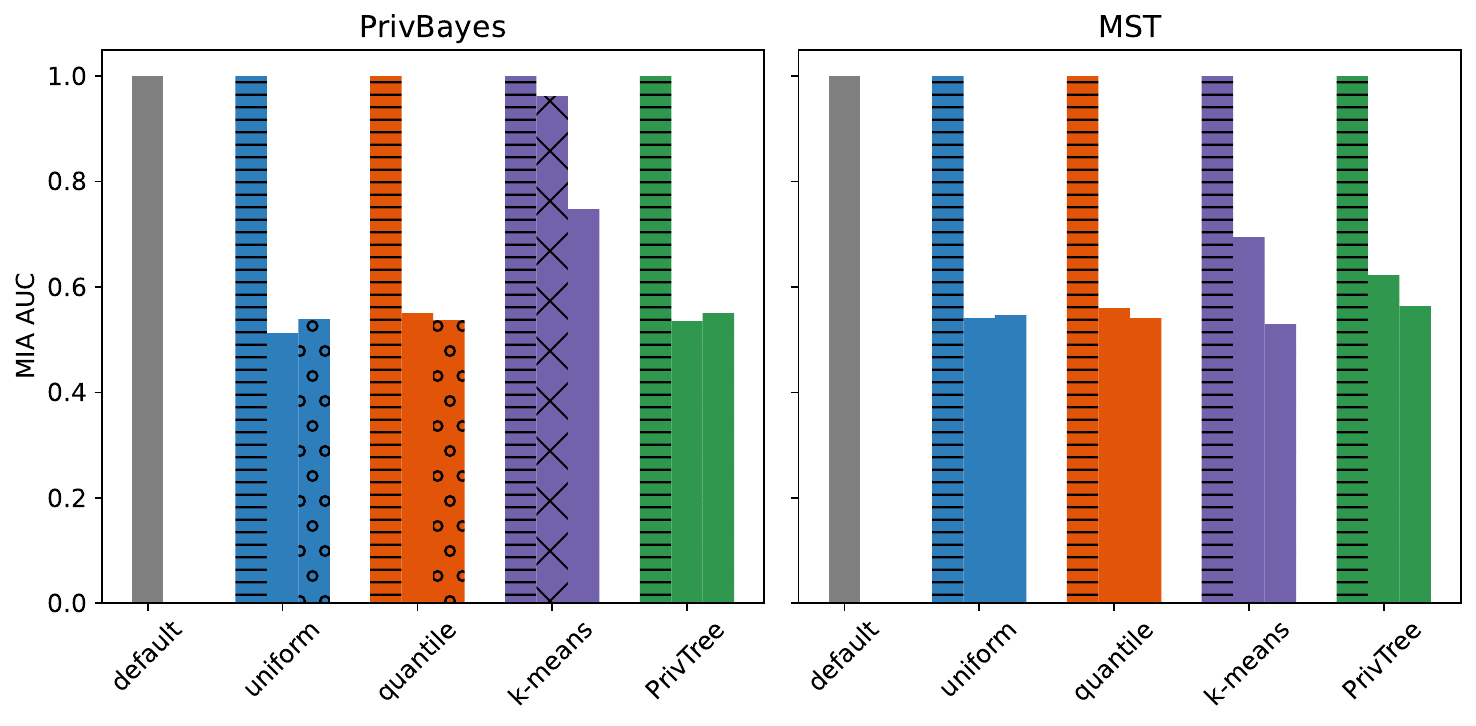}
    \caption{Discretizer ($\epsilon=100$), \\Generator ($\epsilon=1$)}
  \end{subfigure}
\hfill
  \begin{subfigure}{0.3275\linewidth}
    \includegraphics[width=0.99\linewidth]{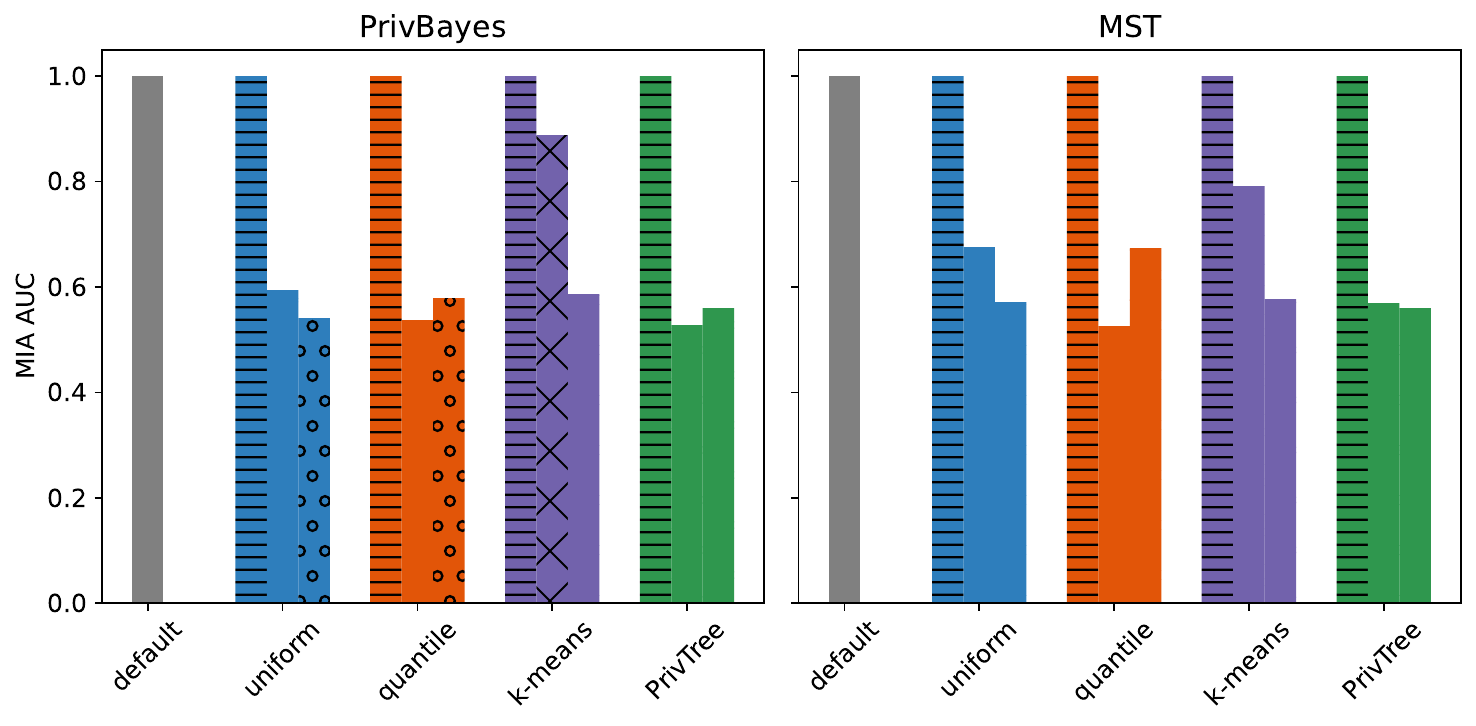}
    \caption{Discretizer ($\epsilon=100$), \\Generator ($\epsilon=100$)}
  \end{subfigure}
  \smallskip
  \caption{\small Privacy leakage with \emph{provided} domain and \emph{extracted} domain (w/ and w/o DP) of the four DP discretizers ($\epsilon=1 \text{ or } 100$) and two DP generative models ($\epsilon=1 \text{ or } 100$) on a target record \emph{outside} the domain of the remaining data.}
  \label{fig:attack_100}
\end{figure*}

\begin{figure*}[t!]
  \centering
  \begin{subfigure}{0.8\linewidth}
    \includegraphics[width=0.99\linewidth]{plots/legend.pdf}
  \end{subfigure}\\
   \begin{subfigure}[t]{0.3275\linewidth}
    \includegraphics[width=0.99\linewidth]{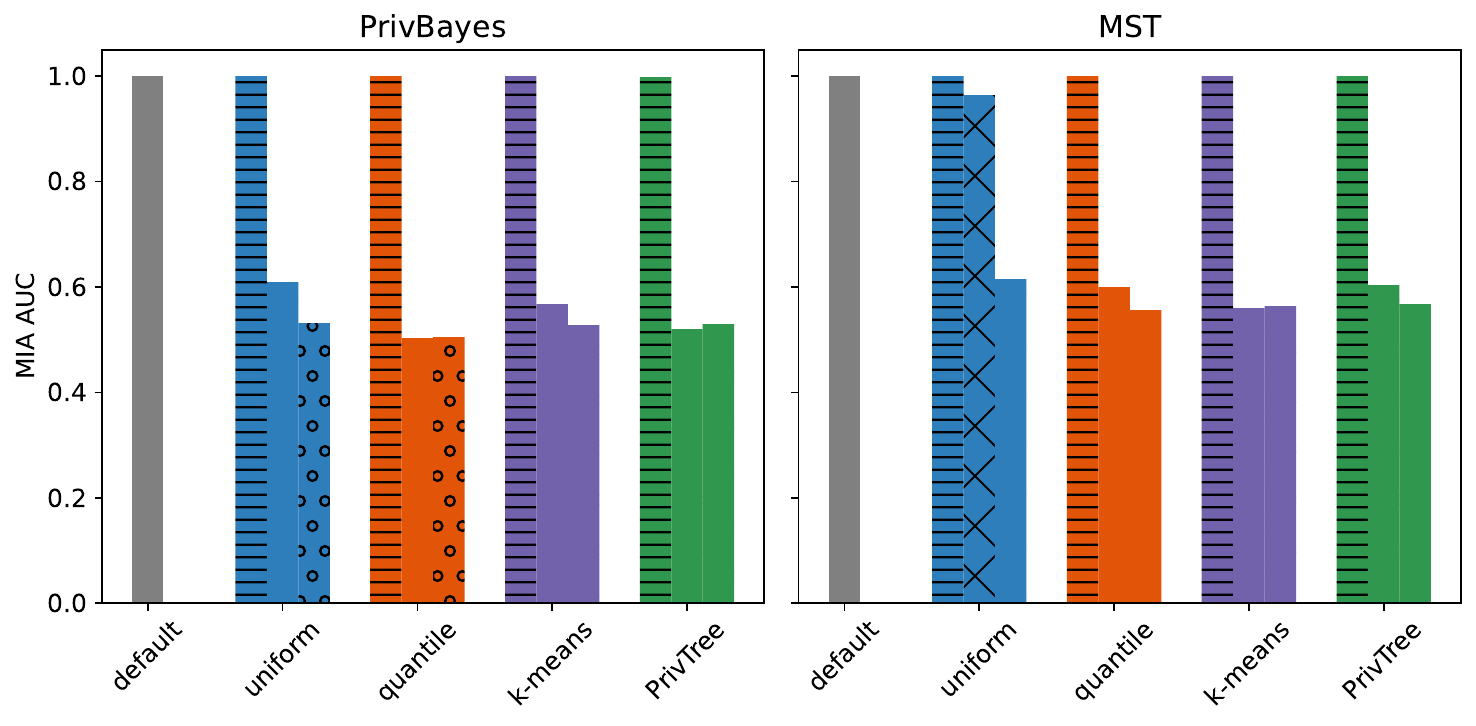}
    \caption{Discretizer ($\epsilon=1$), \\Generator ($\epsilon=1,000$)}
    \label{fig:attack_1_1000}
  \end{subfigure}
\hfill
  \begin{subfigure}[t]{0.3275\linewidth}
    \includegraphics[width=0.99\linewidth]{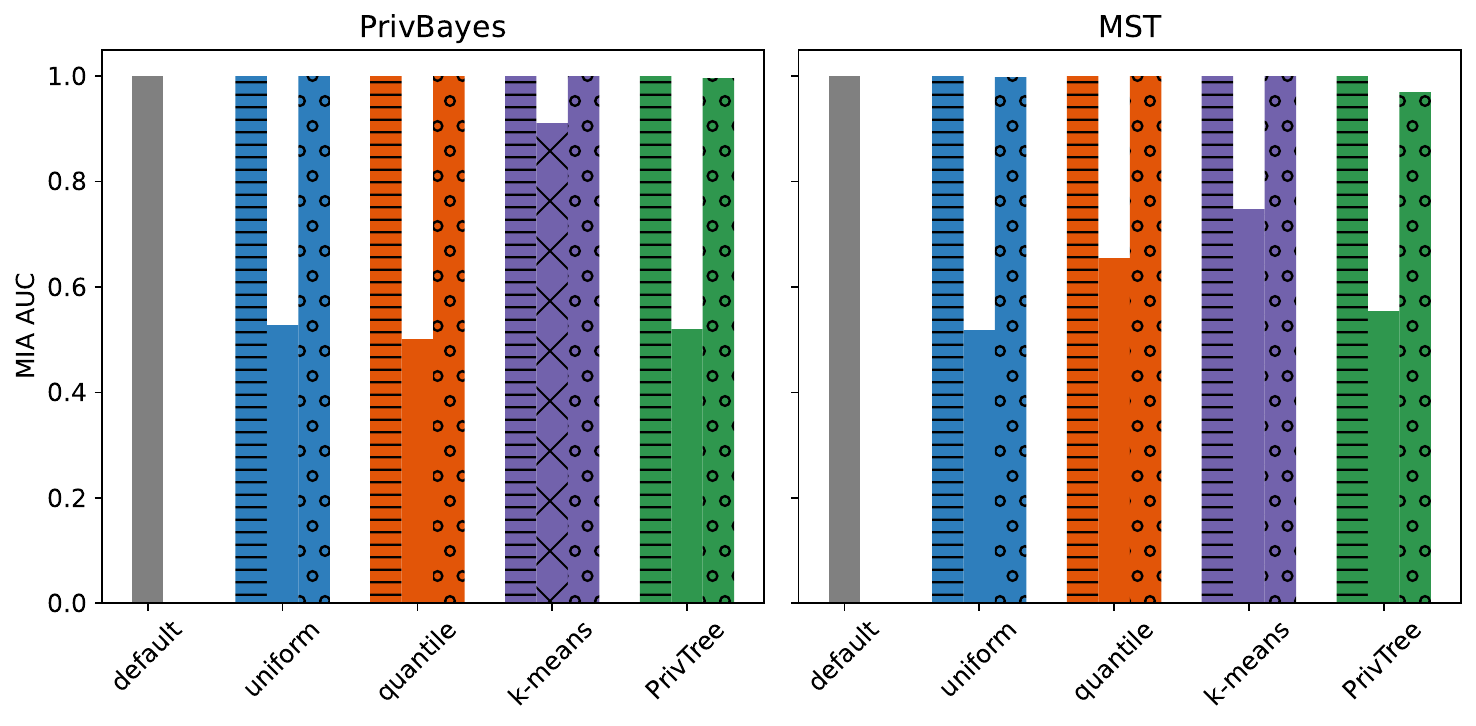}
    \caption{Discretizer ($\epsilon=1,000$), \\Generator ($\epsilon=1$)}
    \label{fig:attack_1000_1}
\hfill
  \end{subfigure}
  \begin{subfigure}[t]{0.3275\linewidth}
    \includegraphics[width=0.99\linewidth]{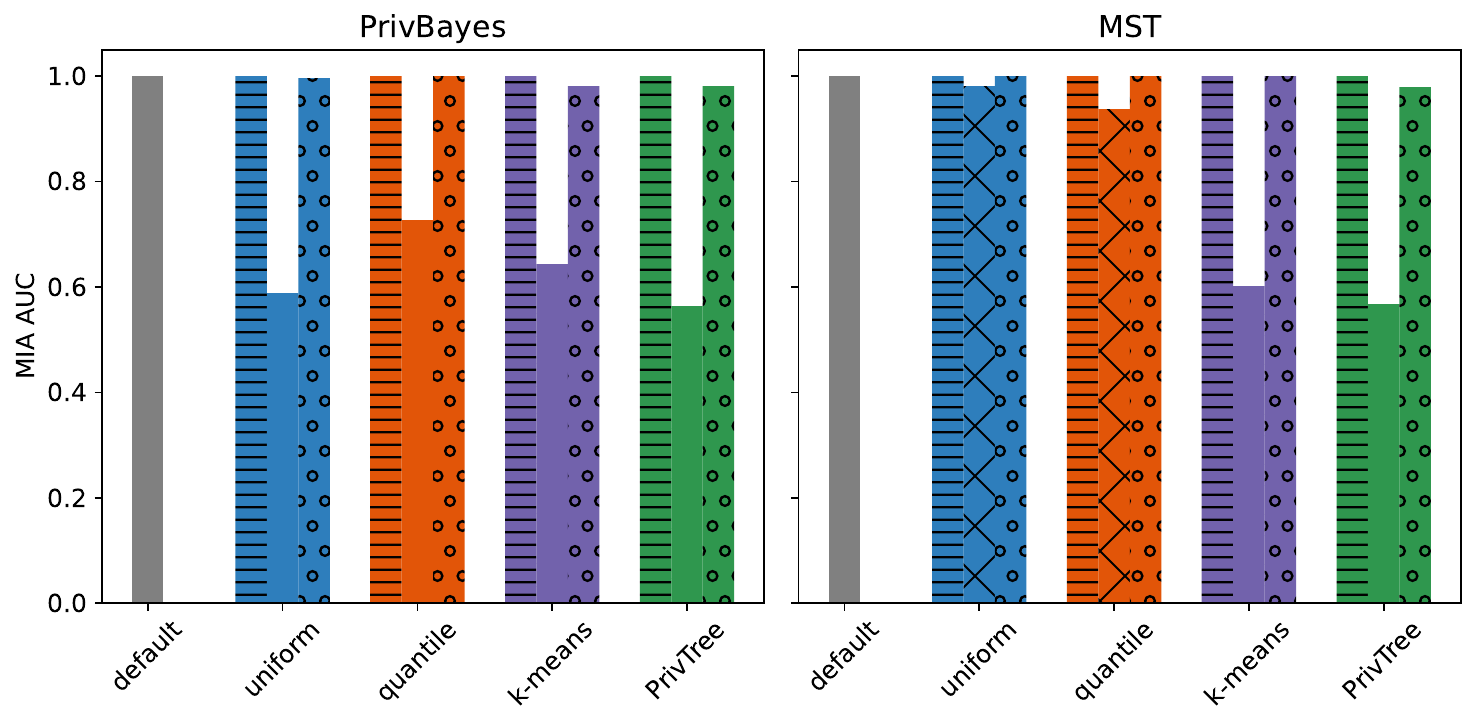}
    \caption{Discretizer ($\epsilon=1,000$), \\Generator ($\epsilon=1,000$)}
    \label{fig:attack_1000_1000}
  \end{subfigure}
\smallskip
  \caption{\small Privacy leakage with \emph{provided} domain and \emph{extracted} domain (w/ and w/o DP) of the four DP discretizers ($\epsilon=1 \text{ or } 1,000$) and two DP generative models ($\epsilon=1 \text{ or } 1,000$) on a target record \emph{outside} the domain of the remaining data.}
  \label{fig:attack_1000}
\end{figure*}

\begin{figure*}[t!]
  \centering
  \begin{subfigure}{0.8\linewidth}
    \includegraphics[width=0.99\linewidth]{plots/legend.pdf}
  \end{subfigure}
  \centering
  \begin{subfigure}{0.3275\linewidth}
    \includegraphics[width=0.99\linewidth]{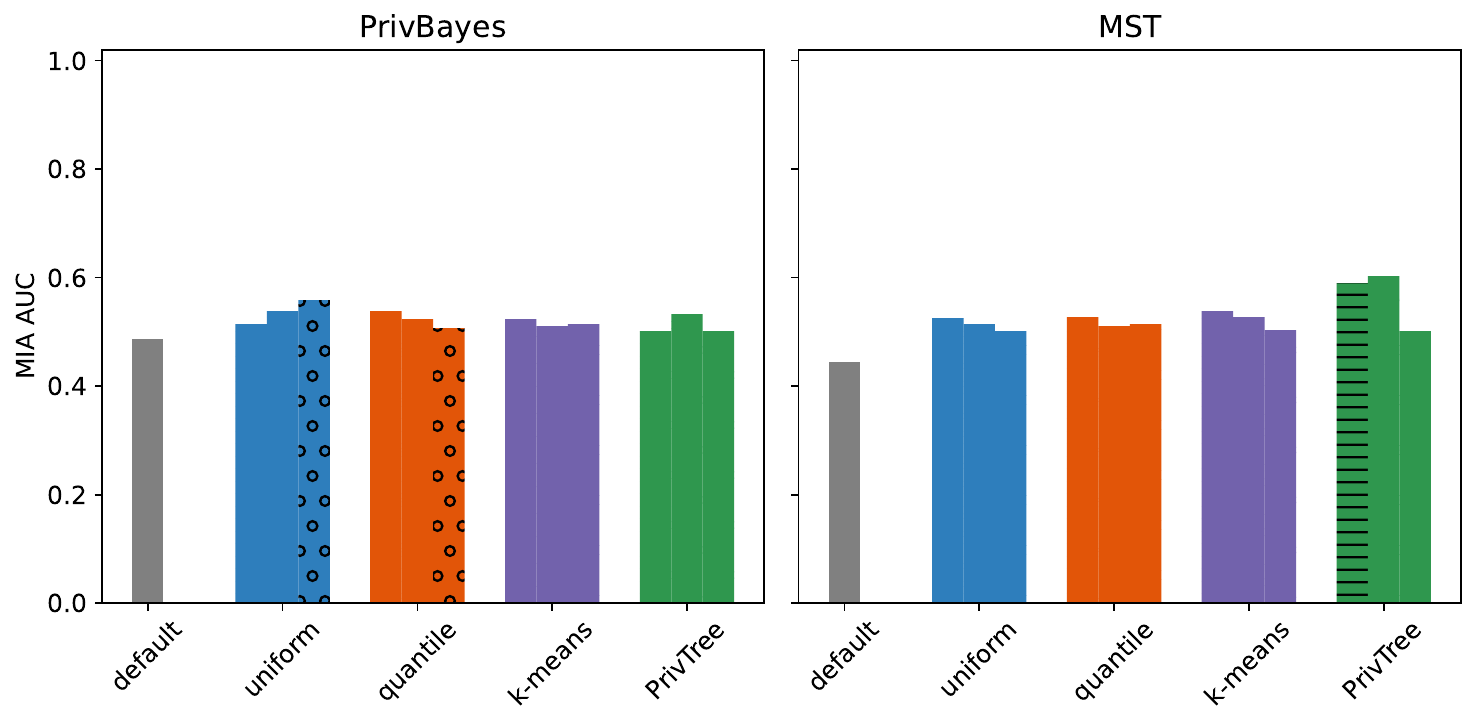}
    \caption{Discretizer ($\epsilon=1$), \\Generator ($\epsilon=1$)}
  \end{subfigure}
\hfill
  \begin{subfigure}{0.3275\linewidth}
    \includegraphics[width=0.99\linewidth]{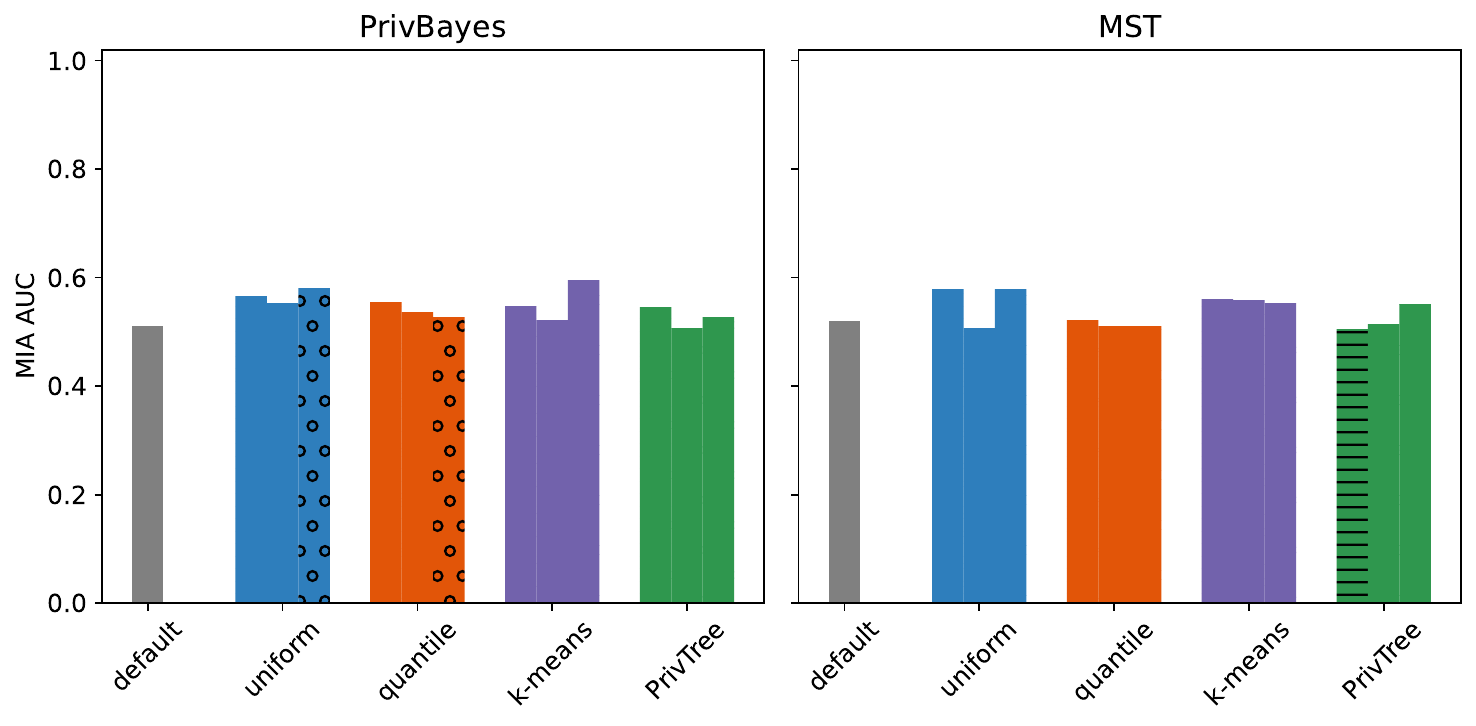}
    \caption{Discretizer ($\epsilon=100$), \\Generator ($\epsilon=100$)}
  \end{subfigure}
\hfill
  \begin{subfigure}{0.3275\linewidth}
    \includegraphics[width=0.99\linewidth]{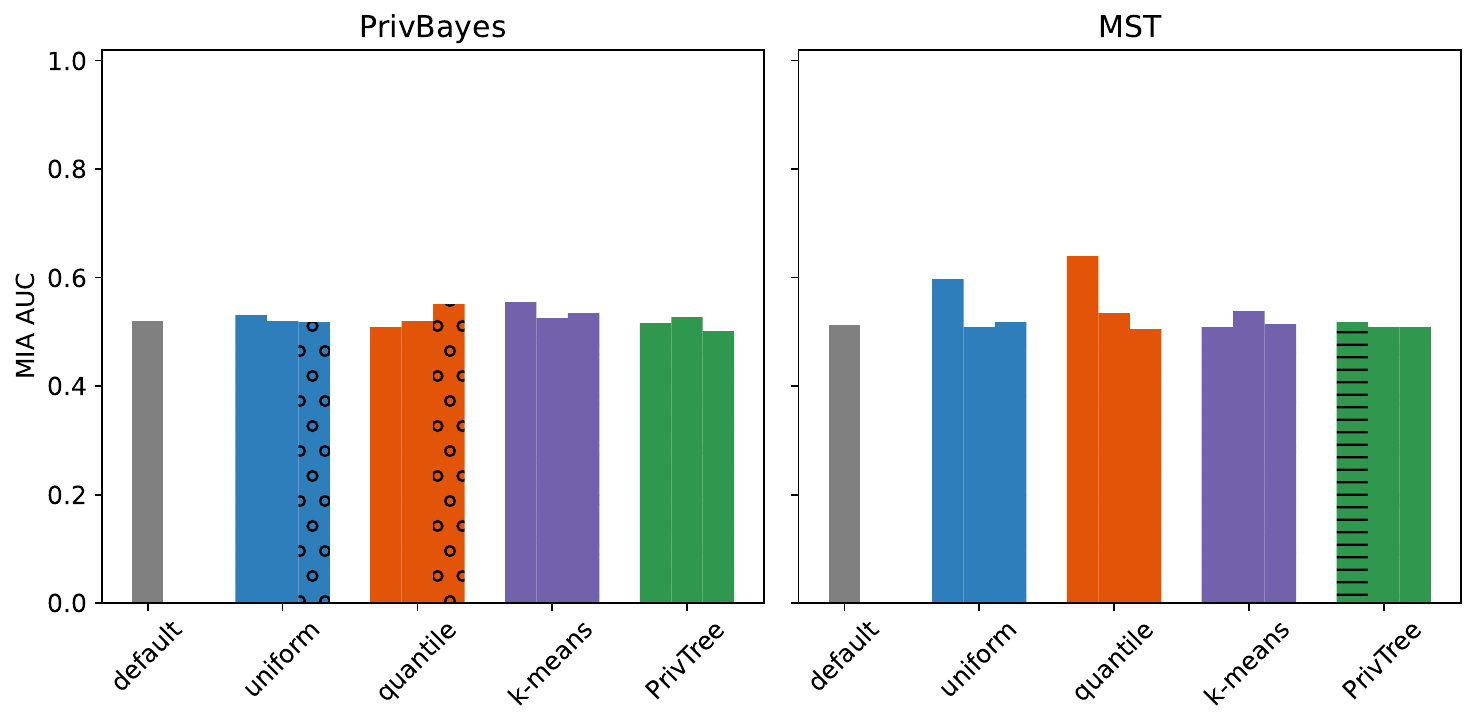}
    \caption{Discretizer ($\epsilon=1,000$), \\Generator ($\epsilon=1,000$)}
  \end{subfigure}
  \smallskip
  \caption{\small Privacy leakage with \emph{provided} domain and \emph{extracted} domain (w/ and w/o DP) of the four DP discretizers ($\epsilon=1, 100 \text{ or } 1,000$) and two DP generative models ($\epsilon=1, 100 \text{ or } 1,000$) on a target record \emph{inside} the domain of the remaining data.}
  \label{fig:attack_in}
\end{figure*}

In Figure~\ref{fig:attack_100} and~\ref{fig:attack_1000}, we present additional results related to running the GroundHog attack~\citep{stadler2022synthetic} on PrivBayes and MST -- specifically, with a target record outside the domain of the remaining data with $\epsilon=1/100 \text{ and } 1/1,000$, respectively.
Figure~\ref{fig:attack_in} also shows results with a target record inside the domain with $\epsilon=1, 100 \text{ and } 1,000$.
We discuss these results in Section~\ref{sec:exp}.

\end{document}